\documentclass[aps,preprint]{revtex4-1}

\usepackage{graphicx}
\usepackage{float}
\usepackage{amsmath}

\begin{document}

\title{Structure and thermodynamics of a mixture of patchy and spherical colloids: a multi-body association theory with complete reference fluid information}
\author{Artee Bansal}
\author{D.\ Asthagiri}
\author{Kenneth R. Cox}
\author{Walter G. Chapman}\thanks{wgchap@rice.edu}
\affiliation{Department of Chemical and Biomolecular Engineering, Rice University, Houston}
\date{\today}
\vfill

\begin{abstract}
A mixture of solvent particles with short-range, directional interactions and solute particles with short-range, isotropic interactions that
can bond multiple times is of fundamental interest in understanding liquids and colloidal mixtures. Because of multi-body correlations 
predicting the structure and thermodynamics of such systems remains a challenge. Earlier Marshall and Chapman developed a theory 
wherein association effects due to interactions multiply the partition function for
clustering of particles in a reference hard-sphere system. The multi-body effects are incorporated in the clustering process, which in their
work was obtained in the absence of the bulk medium. The bulk solvent effects were then modeled approximately within a second order 
perturbation approach. However, their approach is inadequate at high densities and for large association strengths. 
Based on the idea that the clustering of solvent in a defined coordination volume around the solute is related to occupancy statistics
in that defined coordination volume, we develop an approach to incorporate the complete information about hard-sphere clustering 
in a bulk solvent at the density of interest. The occupancy probabilities are obtained from enhanced sampling simulations but we also develop
a concise parametric form to model these probabilities using the quasichemical theory of solutions. We show that incorporating the complete reference 
information results in an approach that can predict the bonding state and thermodynamics of the colloidal solute for a wide range of 
system conditions. 

\end{abstract}

\maketitle

\section{Introduction}
The physical mechanisms governing the structure and dynamics of particles interacting with short-range anisotropic interactions are
of fundamental interest in the quest to understand how inter-molecular interactions dictate macroscopic structural and functional organization   \cite{jackson_phase_1988,marshall_thermodynamic_2014,marshall_cluster_2014,russo_re-entrant_2011,tavares_criticality_2009}. Patchy colloids, particles with engineered directional interactions, are archetypes of such systems, with numerous emerging applications in designing materials from the nanoscale level \cite{glotzer_anisotropy_2007,pawar_fabrication_2010,bianchi_patchy_2011,sciortino_gel-forming_2008,cordier_self-healing_2008}. 
Experiments on patchy colloidal systems have focused on the synthesis of different kinds of self assembling units and their
consequence for the emergent structure \cite{zhang_patterning_2005,yake_site-specific_2007,snyder_nanoscale_2005,chen_directed_2011,pawar_patchy_2008,wang_colloids_2012,romano_colloidal_2011,yi_recent_2013}.  
Complementing these experimental studies, molecular simulations have also sought to understand how the anisotropy of interactions 
determines the emergent structure \cite{zhang_self-assembly_2004,zhang_self-assembly_2005,coluzza_design_2012,de_michele_dynamics_2006} and
the phase behavior \cite{bianchi_phase_2006,bianchi_theoretical_2008,foffi_possibility_2007,giacometti_effects_2010,liu_vapor-liquid_2007,romano_gasliquid_2007}.  But, despite the simplicity in describing and engineering the inter-molecular interactions, a general theory to predict the phase behavior is not yet available. The present article addresses this challenge. 

Wertheim's theory in the form of the statistical associating fluid theory (SAFT) \cite{jackson_phase_1988,wertheim_fluids_1984,chapman_new_1990,wertheim_fluids_1984-1,heras_phase_2011,heras_bicontinuous_2012,bianchi_theoretical_2008,sciortino_self-assembly_2007} has proven effective in describing systems with short range, directional (i.e.\ specific) interactions and is thus of natural interest in describing patchy colloids. In Wertheim's approach, association due to specific interaction is described within a chemical equilibrium framework,  with inter-particle correlations obtained using a non-associating reference fluid (typically a hard-sphere fluid). The nature of information from the reference determines the order of the theory. In the first order (TPT1) theory, pair-correlation information, and in the second order (TPT2) approach, three-body correlation from the reference is included in the theory. 

Wertheim's approach forms the basis of several recent studies that reveal the importance of multi-body effects. Recognizing that patchiness broadens 
the vapor-liquid coexistence relative to a system with isotropic interactions,  Liu et al.\ \cite{liu_vapor-liquid_2007} incorporated a square well reference in 
Wertheim's first order (TPT1) perturbation theory. This model could qualitatively capture the increasing critical temperature with increasing number of 
patches but quantitative accuracy was limited.  Within an integral equation approach, Kalyuzhnyi and Stell \cite{kalyuzhnyi_effects_1993} reformulated Wertheim's multi-density formalism \cite{wertheim_fluids_1986} to incorporate spherically symmetric interactions. However, the solution becomes complex for large values of bonding states. % Wertheim had also developed a second-order (TPT2) perturbation theory\cite{wertheim_thermodynamic_1987} to incorporate the effect of chain and ring formation within the single chain approximation. 
Using TPT2 Phan et al. \cite{phan_equations_1993} developed an equation of state for hard chain molecules and obtained numerical solution for mixtures of chain and star-like molecules. Based on Phan et al.'s\cite{phan_equations_1993} equation for linear chains, Marshall et al.\cite{marshall_three_2013} obtained analytical solutions for branched chain  and star-molecules and their results matched with the numerical results of Phan et al.\cite{phan_equations_1993} for star-like molecules.

To incorporate multi-body effects in SAFT when the association potential of the solute is such as to allow multiple bonding per site, Marshall and Chapman \cite{marshall_molecular_2013,marshall_thermodynamic_2013} extended Wertheim's multi-density formalism to multi-site associating fluids \cite{wertheim_fluids_1986}. This theory generalizes the single chain approximation of Wertheim\cite{wertheim_thermodynamic_1987} for a site bonding multiple times, but 
it requires the multi-body correlation function for solvent around the solute in the non-associating reference fluid. These multi-body correlations for the reference
fluid were sought by characterizing the distribution of gas-phase solute-solvent clusters. The effect of the bulk solvent is subsequently incorporated at the TPT2 level by
using the linear superposition of the pair correlation function plus a three body correction. This approximation works well for systems at low solvent densities or higher concentration of solute, i.e. conditions when low order correlations are important. However, this approach fails at high solvent densities and for low concentrations of solutes. In this work, we address these limitations and present a way to accurately incorporate multi-body correlations in the hard sphere reference.

Following Marshall and Chapman \cite{marshall_molecular_2013,marshall_thermodynamic_2013}, we model  solute and patchy solvent molecules as hard spheres of equal diameter ($\sigma$) and short range association sites. For the hard-sphere reference, we show how coordination distribution around a distinguished solute obtained from particle simulations can be used to incorporate multi-body correlations in the cluster integrals that appear in the theory. The link between coordination distribution and multi-body correlations is inspired by quasichemical theory \cite{pratt_quasichemical_2001,pratt_selfconsistent_2003} and ultimately draws upon Reiss and coworker's  seminal investigation of hard-sphere packing \cite{reiss_statistical_1959}. {In the present framework, all contributions
from orientation-dependent attractive interactions are transparently decoupled from multi-body packing effects that are obtained for the reference fluid.}

The rest of the paper is organized in the following way.  In Section~\ref{sc:bentheory} we discuss the Marshall-Chapman\cite{marshall_thermodynamic_2013} theory, 
their gas-phase cluster approximation and highlight the need for improvement suggested by comparing the results of the theory with Monte Carlo simulations. 
In Section~\ref{sc:correction} our approach for better describing the reference. To aid concision, the development based on the quasichemical theory 
of hard-sphere solutions is presented in the appendix (Section~\ref{sc:qct}). Section~\ref{sc:results} collects the results and discussions of this study.

\section{Theory} 
   
\begin{figure}[h!]
\includegraphics[scale=0.5]{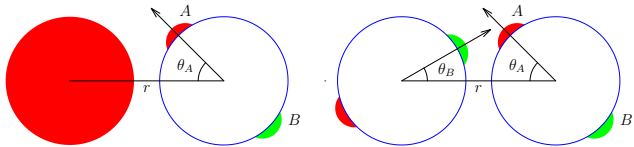}
 \caption{Association between solute and solvent (left) and solvent molecules (right). $r$ is the center-to-center distance and $\theta_A$ and $\theta_B$ are the orientation of the attractive patches $A$ and $B$ relative to line connecting the centers. Note the sticky solute (colored red) can only interact with the patch $A$ (colored red).}\label{fig:1}    
\end{figure}

Consider a mixture of solvent molecules, $p$, with two directional sites (labeled $A$ and $B$) and isotropically-sticky, solute molecules, $s$. 
For solvent-solvent association, only bonding between $A$ and $B$ is allowed and  the size of sites  is such that single bonding condition holds (Fig.~\ref{fig:1}). The solute molecule can bond with site $A$ of the solvent; the isotropic attraction ensures the solute can bond multiple solvent molecules (Fig.~\ref{fig:1}). In the infinitely dilute regime considered here, we ignore the association between the solutes themselves. 

   The association potential for solvent-solvent  $(p,p)$  and solute-solvent $(s,p)$ molecules is given by:
   \begin{equation}
   u_{AB}^{(p,p)}{(r)}=
   \begin{cases}
   -\epsilon_{AB}^{(p,p)}, r<r_c \,\text{and}\, \theta_A\leq \theta_c^{(A)}\,\text{and}\,\theta_B \leq \theta_c^{(B)}
   \\
   0   \text{ \ \ \ \ \ otherwise}
   \\    
   \end{cases}
   \label{eq:1}
   \end{equation}
   
   \begin{equation}
   u_A^{(s,p)}{(r)}=
   \begin{cases}
   -\epsilon_A^{(s,p)}, r<r_c\, \text{and}\, \theta_A\leq \theta_c^{(A)}
   \\
   0   \text{ \ \ \ \ \ otherwise}
   \\    
   \end{cases}
   \label{eq:2}
   \end{equation}
   
   where subscripts $A$ and $B$ represent the type of site and $\epsilon$  is the association energy. $r$ is the distance between the particles and $\theta_A$ is the angle between the vector connecting the centers of two molecules and the vector connecting association site $A$ to the center of that molecule (Fig.~\ref{fig:1}).  The critical distance beyond which particles do not interact is $r_c$ and $\theta_c$ is the solid angle beyond which sites cannot bond. 
   
   In the Wertheim's multi density formalism\cite{wertheim_fluids_1986,wertheim_fluids_1986-1}, the free energy due to association can be expressed as
   \begin{equation}
   	\frac{{A^{AS}}}{{V{k_{\rm B}}T}} = \sum {\left( {{\rho ^{\left( k \right)}}\ln \frac{{\rho _0^{\left( k \right)}}}{{{\rho ^{\left( k \right)}}}} + {Q^{\left( k \right)}} + {\rho ^{\left( k \right)}}} \right)}  - {{\Delta {c^{(0)}}} \mathord{\left/
   			{\vphantom {{\Delta {c^{(0)}}} V}} \right.
   			\kern-\nulldelimiterspace} V}
   	\label{eq:3}
   \end{equation}
   where $k_{\rm B}$ is the
   Boltzmann constant, $T$ is the temperature, the summation is over the species ($k={s,p}$),  $\rho$ is the number density, $\rho_0$ is the monomer density, $Q^{(k)}$ is obtained from Marshall-Chapman\cite{marshall_thermodynamic_2013} work and  $\Delta{c^{(0)}}$ is the contribution to the graph sum due to association between the solvent-solvent  $(p,p)$  and solute-solvent $(s,p)$ molecules, i.e.
   \begin{equation}
   	\Delta c_{}^{\left( 0 \right)} = \Delta c_{pp}^{\left( 0 \right)} + \Delta c_{sp}^{\left( 0 \right)}
   	\label{eq:4}
   \end{equation}
  
 \subsection{Marshall-Chapman theory} \label{sc:bentheory}
\begin{figure*}[ht!]
       	\centering
       	\includegraphics[scale=0.5]{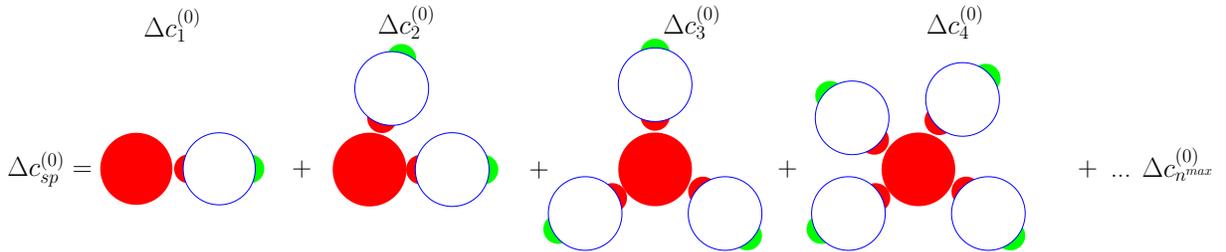}
       	\caption{Representation of graph sums for all the possible arrangements due to association of the solvent around a solute. Note that, for example, the
	graph for $\Delta c_3^{(0)}$ will include other higher occupancy states and is thus  a measure of the effective association with 3 solvent
	particles.}\label{fig:delcn} 
       \end{figure*}

 In the Marshall-Chapman\cite{marshall_molecular_2013,marshall_thermodynamic_2013} work, the role of attractions between solvent, $p$, molecules is accounted by standard first order thermodynamic perturbation theory \cite{jackson_phase_1988} (TPT1). For the association contribution to intermolecular interactions between the  solute ($s$ molecules) and solvent ( $p$ molecules),  Marshall and Chapman \cite{marshall_molecular_2013,marshall_thermodynamic_2013} developed a theory based on generalization of Wertheim's single chain approximation\cite{wertheim_fluids_1986,wertheim_thermodynamic_1987}. By including graph sums for all the possible arrangements of the solvent around the solute i.e.\ one solvent around solute, two solvents around solute, etc. (Fig.~\ref{fig:delcn}),  Marshall and Chapman obtained the free energy expression for the mixture as: 

     \begin{equation}
     	\Delta c_{sp}^{\left( 0 \right)} = \sum\limits_{n = 1}^{{n^{\max }}} {\Delta c_n^{\left( 0 \right)}} 
     	\label{eq:5}
     \end{equation}

 where

\begin{widetext}         
  \begin{eqnarray}
  	\Delta c_n^{(0)} & = & \frac{\rho _0^{(s)} {( {\rho ^{(p)}X_A^{(p)}})}^n} {\tilde \Omega^{n + 1} n!} \int d(1)\cdots d(n + 1) \,
  	g_{HS}( 1 \cdots n + 1) \cdot \prod\limits_{k = 2}^{n + 1} {( f_{A}^{(s,p)} ( 1,k))}  \, .
  	\label{eq:14}
  \end{eqnarray}
\end{widetext}

 In Eq.~\ref{eq:14}, $\rho^{(p)}(=\rho\cdot x^{(p)})$ is the density of solvent molecules obtained from the mole fraction of solvent($x^{(p)}$) and the total density($\rho$), $X_A^{(p)}$ is the fraction of solvent molecules not bonded at site A,  $\tilde \Omega (=4\pi)$ is the total number of orientations, $f_{A}^{( {s,p})}(1,k)( = \exp (\varepsilon_A^{(s,p)}/k_BT) - 1)$  is the  Mayer function for association between $p$ and $s$ molecules corresponding to potential in Eq.~\ref{eq:2} and the integral is over all the orientations and positions of the $n+1$ particles. If the spherical particle is fixed at the origin, the above integral can be represented in terms of the distances from the origin 
\begin{widetext}         
\begin{eqnarray}
\frac{	\Delta c_n^{(0)}}{V} & = & \frac{\rho _0^{(s)} {( {\rho^{(p)}X_A^{(p)}})}^n} {\tilde \Omega^{n} n!} \int d\vec r_1\cdots d\vec r_nd\omega_1\cdots d\omega_n \,
g_{HS}( \vec r_1 \cdots \vec r_n|0) \cdot \prod\limits_{k = 1}^{n} {( f_A^{(s,p)} ( 0,k))}  \, 
\label{eq:n14}
\end{eqnarray}
\end{widetext} 

As is usual in SAFT, the contribution due to association is given by an averaged $f$-bond and factored outside the integral. Integrating over the orientations and defining $\sqrt \kappa_{AA}(=(1-\cos(\theta_c))/2)$ as the probability that molecule $p$ is oriented such that patch $A$ on $p$ bonds to $s$, we get
\begin{widetext}
\begin{eqnarray}
\frac{	\Delta c_n^{(0)}}{V} & = & \frac{\rho _0^{(s)} {( {{\rho}^{(p)}X_A^{(p)}}f_A^{(s,p)}\sqrt {\kappa _{AA}})}^n} {n!} \int_{v} d\vec r_1 \cdots d\vec r_n \,
g_{HS}(\vec r_1 \cdots \vec r_n |0)  \, .
\label{eq:n114}
\end{eqnarray}
\end{widetext}
The limits of the integral in Eq.~\ref{eq:n14} reduce to the volume of the observation shell $(v)$ defined by the region between the diameter of the molecule ($\sigma$) and $r_c$, as the Mayer $f$ function is zero for rest of the positions.  
Due to the limited knowledge of the $m^{th}$ order correlation functions for $m>2$ ($n=1$ corresponds to pair correlation), the calculation of the integral in Eq.~\ref{eq:n114} is a daunting numerical challenge. 

 \subsubsection{Marshall-Chapman approximation (MCA)}\label{sc:MCA}
To simplify numerical calculations, Marshall and Chapman \cite{marshall_molecular_2013} developed an approximation for the cavity 
correlation function, $ {y_{HS}}( \vec r_1 \cdots \vec r_n|0 )$, defined by
 \begin{equation}
     	{g_{HS}}( \vec r_1 \cdots \vec r_n|0) = {y_{HS}}( \vec r_1 \cdots \vec r_n|0 )\prod\limits_{\{ l,k\} } {{e_{HS}} ( {r{}_{lk}})} \, .
     	\label{eq:15}
     \end{equation}
As usual,  $e_{HS}(r_{lk}) = \exp(-u_{HS}(r_{lk})/k_BT) $ are reference system $e$-bonds which serve to prevent hard sphere overlap in the cluster; 
 $e_{HS}(r_{lk}) = 0$ for $r_{lk} = |\vec r_l - \vec r_k| < \sigma$.  At the  TPT2 level, Marshall-Chapman approximated the cavity correlation function
by the first order superposition of pair cavity correlation function at contact corrected by a second order factor
($\delta ^{\left( n \right)}$) to account for three body interactions \cite{marshall_three_2013}, i.e.\ 
   \begin{equation}
   {y_{HS}}(\vec r_1 \cdots \vec r_n|0) \approx y_{HS}^n( \sigma){\delta ^{(n)}} \, .
    \label{eq:yhs}
   \end{equation}

This leads to 
  \begin{eqnarray}
  \int_{v} d\vec r_1 \cdots d\vec r_n\, g_{HS}(\vec r_1 \cdots \vec r_n |0)  \approx y_{HS}^n( \sigma) \delta ^{(n)} \Xi ^{(n)}
\end{eqnarray}
 where
\begin{eqnarray}
 {\Xi ^{(n)}}  =  \int_v d\vec r_1 \int_v d\vec r_2 \ldots \int_v d\vec r_n {\prod\limits_{j > i = 1}^n {e_{HS}(r_{ij} ) } } 
 	\label{eq:110}
\end{eqnarray}

is the partition function for an isolated cluster with $n$ solvent hard spheres around a hard sphere solute in the bonding volume, i.e.\ the spherical shell 
bounded by $\sigma$ and $r_c$, within which particles can associate. 

${\Xi ^{( n )}}$ can be obtained as
   \begin{equation}
   {\Xi ^{( n )}} = \nu _b^n{P^{( n )}}
   \label{eq:11}
   \end{equation}
   where $\nu _b$  is the bonding volume and $P^{( n )}$ is the probability that there is no hard sphere overlap  for randomly generated $p$ molecules in the bonding volume (or inner-shell) of $s$ molecules. A hit-or-miss Monte Carlo \cite{hammersley,pratt_quasichemical_2001} approach to calculate $P^{(n)}$ proves inaccurate for large values of $n$ ($n>8$). But since 
   \begin{equation}
   {P^{( n)}} = P_{insert}^{( n )}{P^{( {n - 1} )}} \, , 
   \label{eq:12}
   \end{equation}
 where $P_{insert}^{(n)}$  is the probability of inserting a {\em{single}} particle given $n-1$ particles are already in the bonding volume, an 
 iterative procedure can be used to build the higher-order partition function from lower order one \cite{marshall_molecular_2013}.   The one-particle insertion probability $P_{insert}^{(n)}$ is easily evaluated using hit-or-miss Monte Carlo.  The maximum number of $p$ molecules for which a non-zero insertion probability can be obtained defines $n^{max}$. 

With the potential defined by Eq.~\ref{eq:2} and the approximation made in Eq.~\ref{eq:yhs}, Eq.~\ref{eq:n114} reduces to
  \begin{equation}
  	\frac{\Delta c_n^{(0)}}{V} = \frac{1}{{n!}}\rho _0^{(s)} \Delta^n {\Xi ^{(n)}}{\delta ^{(n)}}   \,  ,
  	\label{eq:281}
  \end{equation} 
where for a two patch solvent, 
\begin{equation}
\Delta  = {y_{HS}}\left( \sigma \right)X_A^{\left( p \right)}{\rho ^{\left( p \right)}}f_A^{\left( {s,p} \right)}\sqrt {{\kappa _{AA}}} \, .
\label{eq:9}
\end{equation}
The fraction of solute bonded $n$ times is 
  \begin{equation}
  	X_n^{\left( s \right)} = \frac{{\frac{1}{{n!}}{\Delta ^n}{\Xi ^{\left( n \right)}}{\delta ^{\left( n \right)}}}}{{1 + \sum\limits_{n = 1}^{{n^{\max }}} {\frac{1}{{n!}}{\Delta ^n}{\Xi ^{\left( n \right)}}{\delta ^{\left( n \right)}}} }}{{,\quad       n > 0}} \, , 
  	\label{eq:7}
  \end{equation}
and the fraction bonded zero times is
  \begin{equation}
  	X_0^{\left( s \right)} = \frac{1}{{1 + \sum\limits_{n = 1}^{{n^{\max }}} {\frac{1}{{n!}}{\Delta ^n}{\Xi ^{\left( n \right)}}{\delta ^{\left( n \right)}}} }} \, .
  	\label{eq:8}
  \end{equation}

Finally, the average number of solvent associated with the solute is given by: 
   \begin{equation}
   n_{avg} = \sum\limits_n {n\cdot {X^{(s)}_n}}  \, ,
   	\label{eq:81}
   \end{equation}  
    
The fraction of solvent not bonded at site $A$ and site $B$ can be obtained by simultaneous solution of the following equations:
     \begin{equation}
     X_A^{\left( p \right)} = \frac{1}{{1 + \xi {\kappa _{AB}}f_{AB}^{\left( {p,p} \right)}{\rho ^{\left( p \right)}}X_B^{(p)} + \frac{{{\rho ^{\left( s \right)}}}}{{{\rho ^{\left( p \right)}}}}\frac{{ n_{avg} }}{{X_A^{(p)}}}}}  \, ,
     \end{equation}
    \begin{equation}
     X_B^{\left( p \right)} = \frac{1}{{1 + \xi {\kappa _{AB}}f_{AB}^{\left( {p,p} \right)}{\rho ^{\left( p \right)}}X_A^{(p)}}} \, .
     \label{eq:82}
     \end{equation}
  where   
     \begin{eqnarray*}
     \xi  & = & 4\pi {\sigma^2}\left( {{r_c} - \sigma} \right){y_{HS}}(\sigma) \\ 
     \kappa_{AB} & = &\left[1-cos(\theta_c)\right]^2/{4} \\
     f_{AB}^{({p,p})} & = & \exp ( \varepsilon _{AB}^{({p,p})}/k_{\rm B}T)-1 \, .
     \end{eqnarray*}

As will be shown below, the approximation Eq.~\ref{eq:Bn} works very well for low solvent densities ($\rho\sigma^3 < 0.6$), but is inadequate in modeling
a dense system.

 \subsection{The complete reference approach}\label{sc:correction}
\begin{figure*}[ht!]
	\centering
	\includegraphics[scale=0.5]{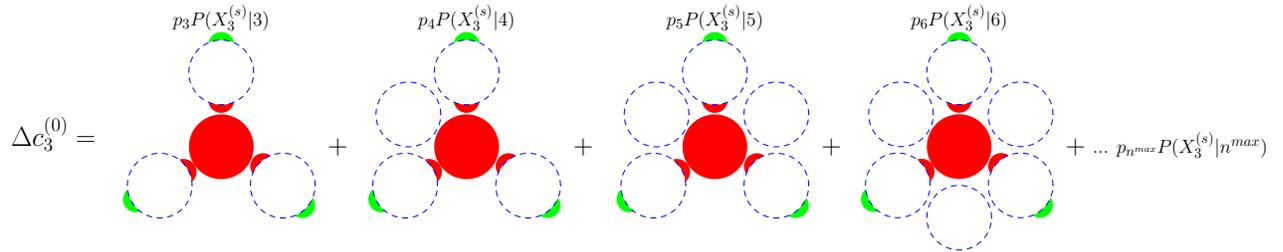}
	\caption{Example graph sum to illustrate the joint role of occupancy and bonding. $p_n$ is the probability of observing $n$-solvent around the solute without regard for their orientation. $P(X_i^{(s)} | n)$ is the conditional    probability that 	given $n$-solvents occupy the bonding volume, $i (\leq n)$ of them are oriented correctly and bond with the solute. Patchy sites for only correctly oriented solvent are shown.  Spheres are indicated by a dashed line to differentiate the graph from the effective association indicated in Fig.~\ref{fig:delcn}. }
	\label{fig:reiss} 
\end{figure*}

The integral appearing in Eq.~\ref{eq:n114} has a simple physical interpretation. It is related to the average number of $n$-solvent clusters (around the distinguished solute) in the hard sphere system \cite{reiss_statistical_1959},  $F^{( n)}$, 
by  
\begin{eqnarray}
 {F^{(n)}} & = &\frac{{{\rho ^n}}}{{n!}}\int\limits_{v} {d{{\vec r}_1} \cdots d{{\vec r}_n}{g_{HS}}\left( {{{\vec r}_1} \cdots {{\vec r}_n}|0} \right)} \nonumber \\
 & = & {\sum\limits_{m = n}^{{n^{\max }}} {C^m_n p_m}} \, ,
 \label{eq:Fn}
 \end{eqnarray}
where $p_n$ is the probability of observing exactly $n$ solvent molecules in the observation shell of the solute in the reference system. $C^m_n$ ($=m! / (m-n)!\cdot n!$) is the combinatorial term which defines the weight for a given coordination state. 
The association contribution (Eq.~\ref{eq:n114}) is then simply 
 \begin{eqnarray}
    \frac{	\Delta c_n^{(0)}}{V}  = {\rho _0^{(s)} {( {x ^{(p)}X_A^{(p)}}f_A^{(s,p)}\sqrt {\kappa _{AA}})}^n} F^{(n)}  \, .
    \label{eq:28}
    \end{eqnarray}
Assuming the availability of $\{p_n\}$, the above approach amounts to including the complete hard-sphere occupancy (packing) information in the
Marshall-Chapman framework. Henceforth, we will refer to this as the ``complete reference" approach.  As figure~\ref{fig:reiss} shows  for
an example of the $\Delta c_3^{(0)}$ term, observe that all occupancy states $m \geq n$ will contribute with combinatorial weights to the bonding state $n$.
Thus errors in accounting for the occupancy of the coordination volume will have a substantial impact in capturing the bonding state. 

Eq.~\ref{eq:28} and the insight derived from it is the central contribution of this work. In support of this, we also present new results 
in modeling of $\{p_n\}$ using ensemble-reweighing approaches and also develop a concise parametric model based on the
quasichemical theory of solutions (Appendix B). As an aside, note that the Marshall-Chapman approximation is
\begin{eqnarray}
{F^{(n)}_{\rm MCA}}  \approx \frac{{{\rho ^n y_{HS}^n( \sigma){\delta ^{(n)}}{\Xi ^{(n)}} }}}{{n!}} \, ,
\label{eq:Bn}
\end{eqnarray} 
   
Given $F^{(n)}$, the fraction of solute associated with $n$ solvent molecules is 
    \begin{equation}
    X_n^{\left( s \right)} = \frac{ {( {x ^{(p)}X_A^{(p)}}f_A^{(s,p)}\sqrt {\kappa _{AA}})}^n F^{(n)} }{{1 + \sum\limits_{n = 1}^{{n^{\max }}} {( {x ^{(p)}X_A^{(p)}}f_A^{(s,p)}\sqrt {\kappa _{AA}})}^n} F^{(n)} } \, ,
    \label{eq:301}
    \end{equation}
and the fraction of with not bonded to any solvent molecule is 
    \begin{equation}
    X_0^{\left( s \right)} = \frac{1}{{1 + \sum\limits_{n = 1}^{{n^{\max }}} {( {x ^{(p)}X_A^{(p)}}f_A^{(s,p)}\sqrt {\kappa _{AA}})}^n} F^{(n)} } \, .
    \label{eq:300}
    \end{equation}
 Using these distributions for associating mixture, average bonded state and fraction of solvent not bonded at sites can be obtained from Eq.~\ref{eq:81}~-~\ref{eq:82}. % 
      
\section{Methods}

\subsection{Monte Carlo simulation of associating system}   
     MC simulations were performed to evaluate the Marshall-Chapman approximation and test the Marshall-Chapman theory with improved representation of the
     multi-body cluster integrals (this work). The associating mixture comprises the sticky solute  and the solvent with 2 diagonally opposed bonding
     sites. For all the simulations, the solvent-solvent and solute-solvent association is defined by the potentials in Eq.~\ref{eq:1} and \ref{eq:2}  respectively, with $r_c = 1.1\sigma$ and  $\theta^{(A)}_c =\theta^{(B)}_c= 27^\circ$.  Unless specifically stated all simulations were based on 255 solvent particles and 1 solute.  
 
 The excess chemical potential of coupling the colloid with the solvent was obtained using thermodynamic
 integration, 
 \begin{eqnarray}
\beta \mu^{Asso} = \epsilon \int_0^1 \langle \beta \psi \rangle_{\epsilon\cdot \lambda} d\lambda \, 
\label{eq:ti}
\end{eqnarray}
where $\langle \beta \psi \rangle_{\epsilon.\lambda}$ is the average binding energy of solute with the solvent as a function of the solute-solvent
interaction strength scaled by $\lambda$ and $\beta = 1 / k_BT$. The integration was performed using a three-point Gauss-Legendre quadrature \cite{Hummer:jcp96}. At each
coupling strength, the system was equilibrated over 1~million sweeps, where a sweep is an attempted move for every particle. 
The translation/rotation factor was chosen to yield an acceptance ratio between $0.3-0.4$. These parameters were kept constant
in the production phase which also extended for 1~million sweeps. Binding strength data was collected every 100 sweeps for analysis. Statistical uncertainty  in $\mu^{Asso} $ was obtained using the Friedberg-Cameron approach \cite{allen:error,friedberg:1970}. 
Simulations were performed at different densities for different interaction schemes and these are specifically noted in the results below. (For $\rho\sigma^3 = 0.9$,
we used 864 particles for better statistics.)
     
To compare the predictions of the bonding state of the colloid ($X_i^{(s)}$) with simulations, from Bayes' rule we have 
\begin{equation}
X_i^{(s)} = \sum_{n\ge i} p_n P(X_i^{(s)} |n) \, ,
\label{eq:bayes}
\end{equation}      
where $P(X_i^{(s)} |n)$ is the probability of observing the colloid in the $i$-bonded state given precisely $n$ solvent particles are in the 
coordination volume. Knowing $X_i^{(s)}$, the average bonding state of the colloid is then 
\begin{equation}
{n_{avg}} = \sum\limits_n n\cdot {X_n^{(s)}}
\label{eq:n_avg}
\end{equation}
To better reveal these low-$X$ states, we used an ensemble reweighting approach \cite{merchant_water_2011}. Biases are calculated
iteratively to sample $n$ as uniformly as possible. The distribution $\{p_n\}$ is readily obtained from the reweighted
probabilities $\{\bar{p}_n\}$ and the converged biases. For each $n$ in the biased simulation, the distribution of $X_i^{(s)}$ is obtained 
and $P(X_i^{(s)}|n)$ constructed. As above, the system was equilibrated over 1~million sweeps and data collected over a production phase of 1~million sweeps. 

To study the effect of concentration of solute on ${n_{avg}}$, it was necessary to explore system with larger number of particles. Specifically, 
we performed simulations for various concentrations  ($0 \leq x_s \leq 1$~) by changing the number of solute molecules in a mixture 
with $864$ number of particles. As above, the system was equilibrated over 1~million sweeps and data collected over a production phase of 1~million sweeps. 
The hard-sphere $\{p_n\}$ distribution was obtained using the reweighting approach \cite{merchant_water_2011}. 

     \subsection{Cluster partition function}
To calculate $P_{insert}^n$ (Eq.~\ref{eq:12}), following Refs.~\onlinecite{marshall_molecular_2013,marshall_thermodynamic_2013}, 
with the solute hard sphere at the center of coordinate system, trial position of the (inserted) solvent in the coordination volume is randomly generated. 
The position is accepted if there is no overlap with either the solute or the remaining $n-1$ particles. The insertion probability is based on similar trial placements
averaged over $10^8-10^9$ insertions. For the present study involving solute and solvent of equal size, the radius of the 
coordination volume is the same as the cut-off radius of $r_c=1.1\sigma$, where $\sigma$ is the hard-sphere diameter.

\section{Results and Discussions}\label{sc:results}
\subsection{Hard Sphere Reference} \label{sc:HSres}     
 \begin{figure*}[h!]
 	\includegraphics[scale=0.8]{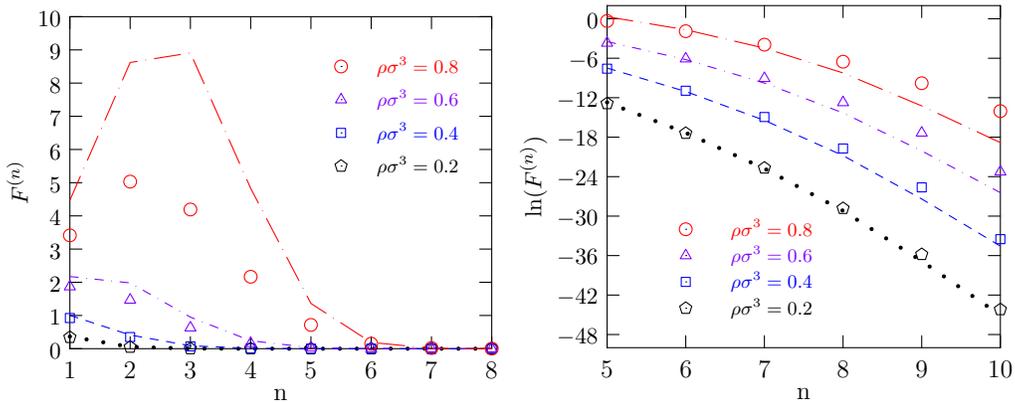}
  	\caption{Comparison of average number of n-mers ($F^{(n)}$) calculated from MC simulation (using Eq.~\ref{eq:Fn}, symbols) and Marshall-Chapman\cite{marshall_molecular_2013} approximation (using Eq.~\ref{eq:Bn}, lines) for different densities (right). n-mers are also presented in natural log scale to clearly show deviation for high coordination states (left). }  
  	\label{fig:n_mers}    
\end{figure*}
Table~ \ref{(table: 1)} (appendix A) collects  the reference $\{p_n\}$ obtained using reweighed sampling for various reduced densities $\rho\sigma^3$.  Figure~\ref{fig:n_mers} shows the comparison of the prediction of $F^{(n)}$ based on the Marshall-Chapman approximation versus molecular
simulations. Observe that for higher solvent densities that are of practical interest in modeling a dense solvent, the Marshall-Chapman 
approximation overestimates the population of lower n-mers and underestimates that of the higher n-mers, but somewhat 
fortuitously it captures n-mers in the range $(6-7)$.  Since the average number of $n$-mer is augmented by $n^{th}$ power of Mayer $f$-function (Eq.~\ref{eq:n114}), the Marshall-Chapman approximation is expected to be progressively inaccurate as the strength of solvent-solute association increases; thus a 
better account of $F^{(n)}$ is needed in securing quantitative accuracy. We next turn to the study of associating mixtures.

\subsection{Associating mixture}

\subsubsection{Solute-solvent versus solvent-solvent association}{\label{sc:asso_bond}}
\begin{figure}[ht!]
	\centering 
	\includegraphics[scale=0.8]{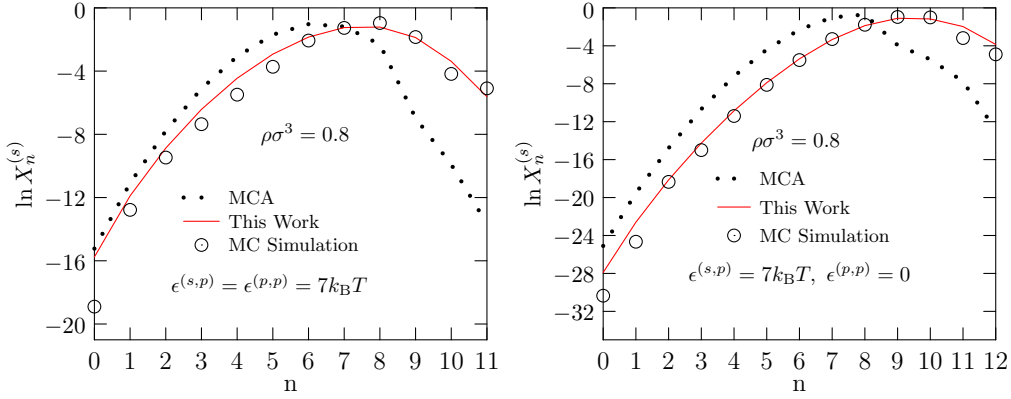}
	\caption{Distribution of bonding states of the solute when the strength of solvent-solvent interaction is 7~$k_{\rm B}T$ (left) and zero (right). The reduced
	density is $\rho\sigma^3=0.8$. $X_n^{(s)}$ is the fraction of solute bonded  $n$-times to the  patchy solvent molecules.  The solute is infinitely dilute and association energy for interaction between solute and solvent is $7k_{\rm B}T$.}
	\label{fig:4}
\end{figure}
Fig.~\ref{fig:4} shows the distribution of the bonding states $\{X_n^{(s)}\}$ of the solute for a reduced density of $\rho \sigma^3=0.8$ for 
two cases, one with and the other without solvent-solvent interaction. In both cases, only one solute is present
in the solvent bath and the solute-solvent interaction is 7~$k_{\rm B}T$.

 Fig.~\ref{fig:4} reveals that for the same interaction energy, higher bonding states are more probable and hence multi-body effects more important in the case 
 when solvent-solvent association is absent. To understand this, note that when both solvent-solvent and solute-solvent interactions are present, there is a competition for patch $A$ on solvent molecule (see Fig.~\ref{fig:1}) to associate with patch $B$ on another solvent molecule or with the sticky solute. This competition  is absent in the case when solvent-solvent association is absent and hence effectively more of the solvent patches are
 available to bond with the solute. Since the Marshall-Chapman approximation has a limitation in capturing the higher $n$-mer state (Fig.~\ref{fig:n_mers}), 
 it is seen that it is unable to capture the distribution of higher bonding states even qualitatively. However, the complete reference approach 
 is able to describe the bonded fraction quite accurately. 

\subsubsection{$\beta \mu^{Asso}$ of solute}
\begin{figure}[ht!]
	\centering
	\includegraphics[scale=0.8]{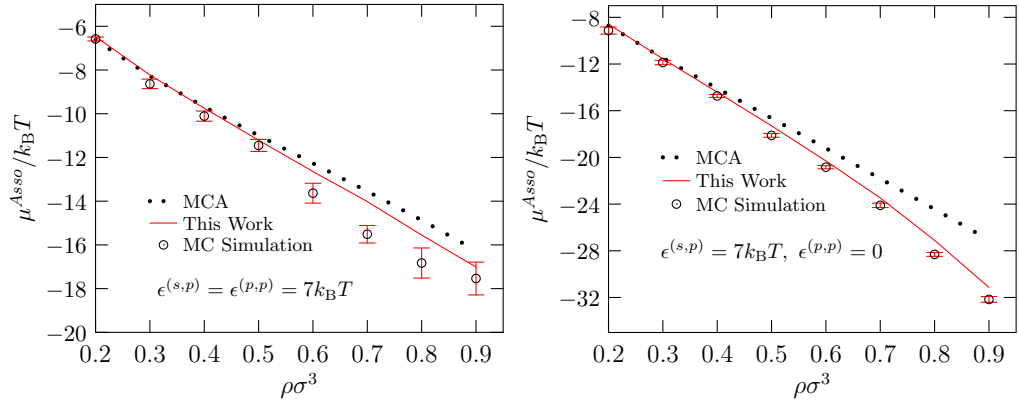}
	\caption{Chemical potential for charging a solute molecule in a patchy solvent environment for different reduced densities. The solute is infinitely dilute and association energy for interactions between solute-solvent molecules is $7k_{\rm B}T$. Solvent-solvent interaction energy is 7~$k_{\rm B}T$(left) and zero (right)}.
	\label{fig:7}
\end{figure}

{\label{sc:muex}}
The ability of the complete reference approach to capture the distribution of bonded states suggests that it should also better describe the association contribution to the
chemical potential. Fig.~\ref{fig:7} supports this suggestion, but for some densities deviations as high as 1~$k_{\rm B}T$ are found. 
Somewhat surprisingly, when solvent-solvent interactions are comparable with solute-solvent interactions,  $\mu^{Asso}$ calculated using
Marshall-Chapman approximation is about as good as the result based with the revised reference (though deviations are larger than the complete reference
approach). 

\begin{figure*}[ht!]
	\centering
	\includegraphics[scale=0.65]{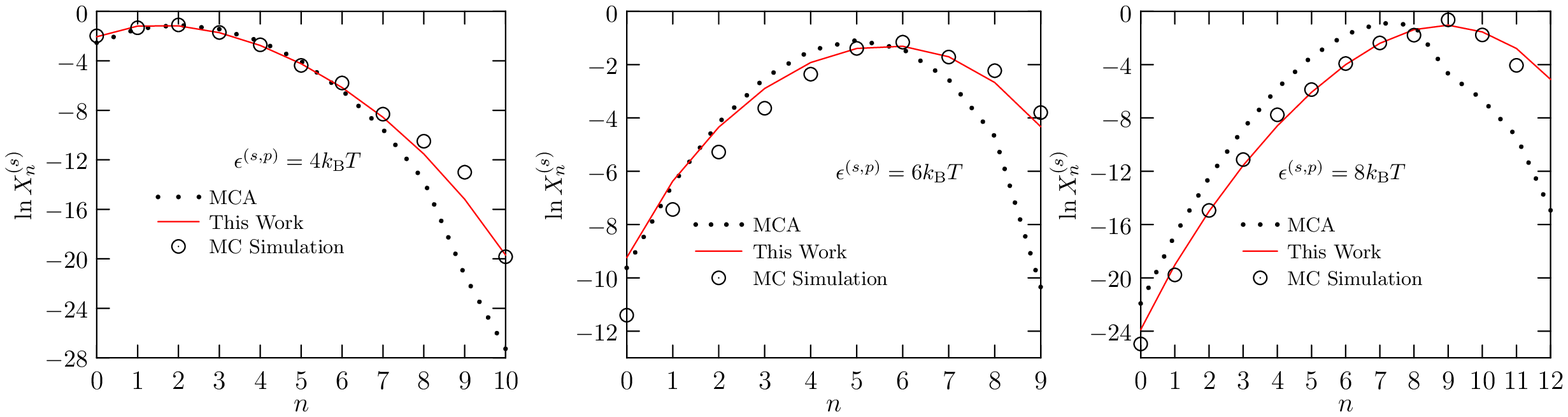}
	\caption{Distribution of bonding states of solute at $\rho\sigma^3=0.8$ and different association energies between solute-solvent molecules; $\epsilon^{(s.p)}=4k_{\rm B}T$ (left),$\epsilon^{(s.p)}=6k_{\rm B}T$ (middle),$\epsilon^{(s.p)}=8k_{\rm B}T$ (right). Solute is infinitely dilute and association energy for solvent-solvent interactions ($\epsilon^{(p,p)}$) is $7k_{\rm B}T$.}
	\label{fig:5_2}
\end{figure*}

Probing the basis for the surprisingly reasonable prediction for $\mu^{Asso}$ based on the Marshall-Chapman approximation reveals the importance
of competitive solute-solvent and solvent-solvent interactions.   When solvent-solvent association is comparable to 
solute-solvent association, specifically, $\epsilon^{(p,p)} \ge \epsilon^{(s,p)}$, the Marshall-Chapman approximation is able to capture the 
bonding states up to the most probable bonding state  reasonably well (Fig.\ref{fig:5_2}). However, when solute-solvent association 
strength is higher than solvent-solvent association, the Marshall-Chapman approximation is unable to capture neither the most probable bonding state nor
the lower bonding states. Since in calculating $\mu^{Asso}$ we integrate the mean binding energy (Eq.~\ref{eq:ti}) over the regime
where $\epsilon^{(p,p)} > \epsilon^{(s,p)}$, and since this is also the regime in which the Marshall-Chapman approximation is comparable to the 
revised theory, the final observed differences in the prediction of the chemical potential are not thus substantial. 

We suspect that the entropic and enthalpic components of $\mu^{Asso}$ will be more sensitive to the description of the reference, an aspect that we are currently investigating. However, the chemical potential results do suggest a cautionary note, namely that a metric based on $\mu^{Asso}$ may mask 
differences in underlying approximations.

\subsubsection{Variation of average bonding with association energy}

\begin{figure}[h!]
 	\centering
 	\includegraphics[scale=0.8]{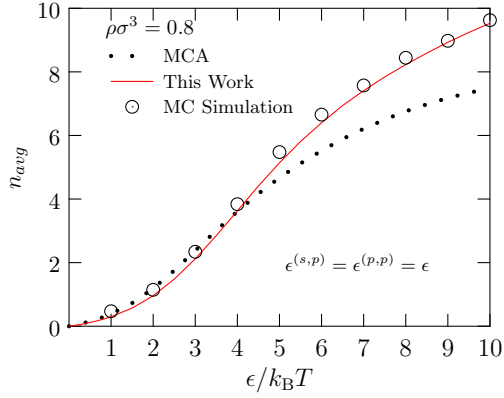}
 	\caption{Variation of average bonding number ($n_{avg}$) with the association energy between molecules $(\epsilon)$  for $\rho\sigma^3=0.8$. Solute is infinitely dilute.}
 	\label{fig:5_1}
 \end{figure}

  Fig.~\ref{fig:5_1} shows the variation of average bonding numbers $(n_{avg})$ for the solute, when the solute-solvent and solvent-solvent 
  association strengths are the same. 
 
It can be observed from Eq.~\ref{eq:Fn} that at high densities,  the contribution to $F^{(n)}$ from the higher-occupancy (higher $p_n$) state is non-negligible. For the association contribution, recall that $F^{(n)}$ is multiplied by $n$-factors of the Mayer $f$ function which itself depends exponentially on strength of association. 
Thus for high density and high association strength, the TPT2 (Marshall-Chapman approximation) prediction is expected to underestimate $n_{avg}$. We find
that this is indeed the case, but using the complete reference approach we can capture $n_{avg}$ accurately. 
\begin{figure}[h!]
	\centering
	\includegraphics[scale=0.8]{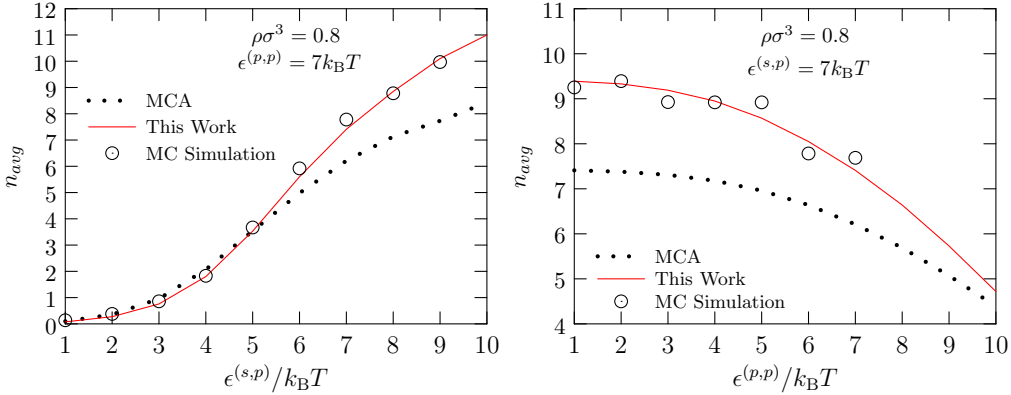}
	\caption{Variation of average bonding number ($n_{avg}$) with the association energy between solute-solvent molecules($\epsilon^{(p,p)}$)(left) and solvent-solvent molecules ($\epsilon^{(s,p)}$)(right) for $\rho\sigma^3=0.8$. Solute is infinitely dilute.}
	\label{fig:5}
\end{figure} 

Fig.\ref{fig:5} further highlights the importance of the relative strengths of solute-solvent and solvent-solvent interactions on multi-body effects.  
As expected from the foregoing analysis, when the solute-solvent interactions are much stronger than solvent-solvent interactions, 
$n_{avg}$ based on the Marshall-Chapman approximation deviates significantly  from $n_{avg}$ based on simulations. 

\subsubsection{Solute concentration effect}
\begin{figure}[h]
 	\centering
 	\includegraphics[scale=0.8]{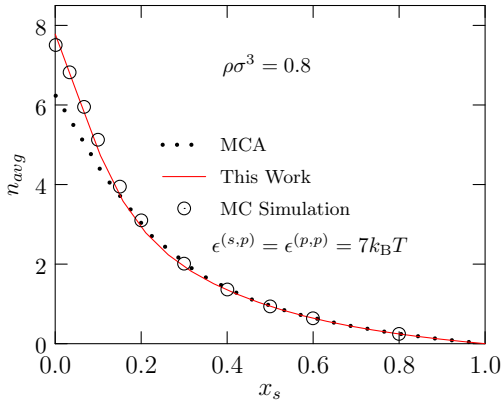}
 	\caption{Variation of average bonding number ($n_{avg}$) with the concentration of solute in the solution for $\rho\sigma^3=0.8$. Energy of association between molecules is $7k_{\rm B}T$.}
 	\label{fig:6}
 	
 \end{figure}

We also study the variation of average bonding number ($n_{avg}$) with the concentration of solute molecules ($x_s$) for a reduced density of $\rho \sigma^3=0.8$. (The solvent-solvent and solute-solvent interactions are all at the same level, $\epsilon=7k_{\rm B}T$). 
For low concentration of solute, since more solvent molecules are available to associate with the solute, higher bonding states are more probable. It is precisely in this limit that we expect larger deviations from the Marshall-Chapman approximation. 
As the concentration of solute is increased, proportionately fewer solvent molecules are available to bond with the colloid. In this limit, 
the effect of multi-body effects in solvation of the colloid should be tempered and better agreement with the Marshall-Chapman approximation is 
expected. Fig.~\ref{fig:6} confirms these expectations.  

 \section{Conclusion}\label{sc:conclusions}
            
In this study we have developed a simple and effective way to model multi-body effects in colloidal mixtures. Building on the Marshall and Chapman theory, we show that 
the challenge in describing the multi-body effects in associating mixtures can be handled by appreciating the importance of packing in the reference system. Importantly, 
we establish that the complex multi-body effects in the associating mixtures of different association geometries can be accurately determined if correct reference information is used.  The present approach can elucidate the structure and thermodynamics of mixtures of patchy-solvent and sticky-solutes with size and interaction asymmetry as well as short-range ion-association phenomena in a dipolar solvent, cases where multi-body effects are potentially important. 
   
    In this work, we  incorporate complete information from the hard-sphere reference fluid and present a modified expression for calculation of associative contribution to graph sums within the framework provided by Marshall-Chapman theory\cite{marshall_molecular_2013,marshall_thermodynamic_2013}.This modified expression is based on the analysis of physical clusters in the hard sphere and their representation in terms of occupancy distribution around a distinguished solute
 in the reference fluid. These occupancy distributions were obtained from enhanced sampling methods for hard sphere systems at different densities. 
Analysis of a wide range of association and concentration regimes shows that our approach incorporating complete hard sphere information 
accurately captures the behavior for bonding states, and the prediction of the chemical potential contribution due to association is within 1~$k_{\rm B}T$ of the
reference Monte Carlo simulation results.

   \section{Acknowledgment} 
We thank Ben Marshall for helpful discussions.  We acknowledge RPSEA / DOE 10121-4204-01 and the Robert A. Welch Foundation (C-1241) for financial support.

 \newpage    

 \section{Appendix} 
% \setcounter{section}{0}
%\makeatletter
%\renewcommand{\thesection}{S.\@Roman\c@section}

\setcounter{figure}{0}
\makeatletter
\renewcommand{\thefigure}{S\@arabic\c@figure}

 \subsection{Coordination number distribution}

\setcounter{table}{0}
\makeatletter
\renewcommand{\thetable}{A.\@Roman\c@table}
 
 \begin{table}[ht]
	\caption{ $\ln(p_n)$ in hard sphere reference system obtained by reweighted sampling for different reduced densities ($\rho \sigma^3$) }  
	\centering
	\begin{tabular}{|c| c| c| c| c|}
		\hline\hline
		$n$ &$\rho \sigma^3=0.6$& $\rho \sigma^3=0.7$ & $\rho \sigma^3=0.8$&$\rho \sigma^3=0.9$\\
		\hline
		0	&	-2.17	&	-3.22	&	-4.74	&	-6.74	\\
		1	&	-1.19	&	-1.78	&	-2.75	&	-4.29	\\
		2	&	-1.15	&	-1.22	&	-1.70	&	-2.75	\\
		3	&	-1.67	&	-1.28	&	-1.27	&	-1.79	\\
		4	&	-2.77	&	-1.87	&	-1.35	&	-1.33	\\
		5	&	-4.44	&	-2.98	&	-1.93	&	-1.35	\\
		6	&	-6.66	&	-4.60	&	-3.05	&	-1.84	\\
		7	&	-9.47	&	-6.85	&	-4.66	&	-2.83	\\
		8	&	-13.20	&	-9.80	&	-6.96	&	-4.39	\\
		9	&	-17.73	&	-13.62	&	-10.06	&	-6.71	\\
		10	&	-23.43	&	-18.64	&	-14.13	&	-9.79	\\
		11	&	-35.23	&	-25.35	&	-18.95	&	    	\\
		12	&	-47.53	&	-32.85	&	-24.88	&	    	\\
		
		\hline
	\end{tabular}
	\label{(table: 1)}
\end{table}

  \subsection{Quasi-chemical theory for solvation of hard-sphere reference}\label{sc:qct}  

\setcounter{table}{0}
\makeatletter
\renewcommand{\thetable}{B.\@Roman\c@table}
 
\setcounter{equation}{0}
\makeatletter
\renewcommand{\theequation}{B.\@arabic\c@equation}

  Consider the equilibrium clustering reaction within some defined coordination volume of the solute $A$ in a bath of solvent $S$ molecules
  \begin{equation}
  A{S_{n = 0}} + {S_n} \rightleftharpoons A{S_n} \, . 
  \label{eq:20}
  \end{equation}
  The equilibrium constant  is
  \begin{equation}
  {K_n} = \frac{{{\rho _{A{S_n}}}}}{{{\rho _{A{S_{n = 0}}}}\rho _s^n}} \, ,
  \label{eq:21}
  \end{equation}
  where $\rho_{AS_n}$ is the density of  species $AS_n$ and $\rho_s$ is the density of the solvent. A mass balance
  then gives the fraction of $n$-coordinated solute as    
  \begin{equation}
  {p_n} = \frac{{{K_n}^{}\rho _s^n}}{{1 + \sum\limits_{m \ge 1} {{K_m}^{}\rho _s^m} }} \, .
  \label{eq:16}
  \end{equation}
  The $n=0$ term, $p_0$, is of special interest: $\ln p_0$ is free energy of allowing solvent molecules to populate a formerly empty coordination shell.   
  In the language of quasichemical theory, $\ln p_0$ is called the chemical term \cite{lrp:book,lrp:cpms,merchant_thermodynamically_2009}. Because the bulk medium pushes solvent into the coordination volume, an effective attraction exists between the solute and solvent even 
  for a hard-sphere reference.

  In the primitive quasichemical approximation \cite{merchant:jcp11b}, the equilibrium constants are evaluated by neglecting the effect of the bulk medium, i.e.\ for an isolated cluster. Thus $K_n \approx K_n^{(0)}$ \cite{pratt_quasichemical_2001}, where 
  \begin{equation}
  n!K_n^{(0)} = \int\limits_v {d {\vec r}_1 \cdots  \int\limits_v d{{\vec r}_n} {\prod\limits_{j > i = 1}^n {e(i,j)} } } \,
  \label{eq:210}
  \end{equation}
with the integration restricted to the coordination volume ($v$).  Comparing  Eqs.~\ref{eq:210} and~\ref{eq:110}, 
  clearly  $n! {K_n}^{(0)} \equiv \Xi ^{(n)}$, establishing a physical meaning for Eq.~\ref{eq:110}. 
  
  It is known that the primitive approximation leading to Eq.~\ref{eq:210} introduces errors in the estimation of $\ln p_0$ \cite{pratt_quasichemical_2001,pratt_selfconsistent_2003}, especially for systems where the interaction of the solute with the solvent is not sufficiently stronger than the interaction amongst solvent particles \cite{merchant:jcp11b}. For hard spheres we must then expect the primitive approximation to fail outside the limit of low 
  solvent densities.

  One approach to improve the primitive approximation is to include an activity coefficient $\zeta_1$, such that the predicted occupancy
  in the observation volume is equal to occupancy, $\langle n\rangle$, expected in the dense reference \cite{pratt_quasichemical_2001}
  \begin{eqnarray}
  \sum\limits_n {n{K_n}^{\left( 0 \right)}\rho _S^n{\zeta_1 ^n}}  = \left\langle n \right\rangle \sum\limits_n {{K_n}^{\left( 0 \right)}\rho _S^n{\zeta_1 ^n}} \, .
  \label{eq:22}
  \end{eqnarray}
  Here the factor $\zeta_1$ functions as a Lagrange multiplier to enforce the required occupancy constraint ($\langle n\rangle$). Physically, 
  $\zeta$ is an activity coefficient that serves to augment the solvent density in the observation volume over that predicted by the gas-phase
  equilibrium constant $K_n^{(0)}$.  
  
Pratt and Ashbaugh \cite{pratt_selfconsistent_2003} showed that $\zeta_1$ alone is inadequate in modeling hard-sphere packing at high
densities, but in addition one needs a solvent coordinate-dependent molecular field to enforce uniformity of density inside the observation volume. 
With this additional molecular field, they showed that using the few-body cluster integrals (Eq.~\ref{eq:22}), they could predict 
hard-sphere packing in excellent agreement with the Carnahan-Starling \cite{mansoori_equilibrium_1971} result up to high 
  densities. Drawing inspiration from the Pratt and Ashbaugh \cite{pratt_selfconsistent_2003} work, 
  we find that a two-parameter model
\begin{equation}
  {p_n} = \frac{ {{K_n}^{(0)}[{{\zeta_1. \exp(\zeta_2. n)}}\rho _S]^n}}{{1 + \sum\limits_{m \ge 1} {{K_m}^{(0)}[{{\zeta_1. \exp(\zeta_2. m)}}\rho _S]^m} }} \ 
\label{eq:pn_new}
\end{equation}
with parameters reported in Table~\ref{(table: Lagrange values)} is able to accurately reproduce both the free energy to evacuate the coordination volume around the reference solute and the average occupancy.
  \begin{table}[ht]
  	\caption{Parameters for Eq.~\ref{eq:pn_new} for different $\rho \sigma^3$ }  
  	\begin{tabular}{|c| c| c|}
  		\hline
  		$\rho \sigma^3$ & $\zeta_1$  & $\zeta_2$  \\
  		\hline
  		0.2 &	1.2991 &	0.0175 \\
  		0.6 &	3.1475 &	0.0361 \\
  		0.7 &	4.1072 &	0.0457 \\
  		0.8	&  5.5875 & 0.0609 \\ 
  		0.9	&   7.5149 & 0.0829 \\
  		\hline
  	\end{tabular}
  	\label{(table: Lagrange values)}
  \end{table}   
  
Interestingly, Eq.~\ref{eq:pn_new} is identical in form with the model Reiss et. al.\cite{reiss_upper_1981} derive to describe the effects of clustering, medium and surface interactions simultaneously in hard-sphere packing in a cavity. Eq.~\ref{eq:pn_new} can also be derived using a MaxEnt procedure with the mean and variance of the occupancy
as constraints. 
  
Eq.~\ref{eq:pn_new} can be used to obtain the average number of $n$-mer (Eq.~\ref{eq:Fn}) in the coordination volume. It was also observed that based on the geometric effects involved in the surface interactions, $\{p_n\}$ can be determined by just one density dependent parameter along with a chemical potential. 
A detailed development of this idea will be presented later.

\newpage

% \bibliography{references} 


\end{document}